\documentclass[twocolumn,aps,prl,showpacs,superscriptaddress,longbibliography,10pt]{revtex4-2} %
\usepackage[colorlinks=true,linkcolor=blue,citecolor=blue]{hyperref}
\usepackage{amsfonts}
\usepackage{subfigure}
\usepackage{amsmath}
\usepackage{mathrsfs}
\usepackage{amssymb}
\usepackage{amsbsy}
\usepackage{epsfig}
\usepackage{graphicx}
\usepackage{epstopdf}
\usepackage{mathdots}
\usepackage{color}
\usepackage{cleveref}
\usepackage{float}
\usepackage{graphicx}
\usepackage{rotating}
\usepackage{physics}
\usepackage{nicefrac}

\begin{document}
\title{Entanglement signatures of quantum criticality in Floquet non-Hermitian topological systems}
\author{Siyuan Cheng}
\affiliation{Department of Physics, Jiangsu University, Zhenjiang, 212013, China}  

\author{Rui Xie}
\affiliation{Department of Physics, Jiangsu University, Zhenjiang, 212013, China}  

\author{Xiaosen Yang}
\affiliation{Department of Physics, Jiangsu University, Zhenjiang, 212013, China}  

\author{Yuee Xie} \altaffiliation{yueexie@ujs.edu.cn}
\affiliation{Department of Physics, Jiangsu University, Zhenjiang, 212013, China}  

\author{Yuanping Chen} \altaffiliation{chenyp@ujs.edu.cn}
\affiliation{Department of Physics, Jiangsu University, Zhenjiang, 212013, China}  

\date{\today}

\begin{abstract}
The entanglement entropy can be an effective diagnostic tool for probing topological phase transitions. In one-dimensional single particle systems, the periodic driving generates a variety of topological phases and edge modes. In this work, we investigate the topological phase transition of the one-dimensional Floquet Su-Schrieffer-Heeger model using entanglement entropy, and construct the phase diagram based on entanglement entropy. The entanglement entropy exhibits pronounced peaks and follows the logarithmic scaling law at the phase transition points, from which we extract the central charge $c=1$. We further investigate the entanglement spectrum to accurately distinguish the different topological phases. In addition, the coupling between zero and $\pi$ modes leads to characteristic splittings in the entanglement spectrum, signaling their hybridization under periodic driving. These results remain robust in non-Hermitian regimes and in the presence of next-nearest-neighbor hopping, demonstrating the reliability and universality of entanglement entropy as a diagnostic for topological phase transitions. 

\end{abstract}
\maketitle

\section{I. Introduction} 
Entanglement entropy is a fundamental concept in modern physics, playing a crucial role in quantum information, quantum gravity, statistical physics, and condensed matter physics \cite{amicoEntanglementManybodySystems2008,nishiokaEntanglementEntropyHolography2018,abaninColloquiumManybodyLocalization2019,zhuIdentifyingCnsymmetricHigherorder2020,kokailEntanglementHamiltonianTomography2021}. In complex materials and systems, entanglement entropy has become a powerful tool for probing characteristics of quantum correlations and critical behavior \cite{kitaevTopologicalEntanglementEntropy2006,levinDetectingTopologicalOrder2006,plenioEntropyEntanglementArea2005,sirkerBoundaryFidelityEntanglement2014,hastingsEntropyEntanglementQuantum2007}. Entanglement entropy characterizes the quantum correlations between subsystems, exhibiting area-law scaling in one-dimensional (1D) gapped systems and logarithmic scaling in 1D gapless systems \cite{korepinUniversalityEntropyScaling2004,wolfViolationEntropicArea2006,leschkeScalingRenyiEntanglement2014,barthelEntanglementScalingCritical2006,liScalingBehaviorEntanglement2006,eisertColloquiumAreaLaws2010,joshiExploringLargescaleEntanglement2023,calabreseEntanglementEntropyOneDimensional2011,dingEntanglementEntropyFermi2012,laflorencieQuantumEntanglementCondensed2016,pasqualecalabreseEntanglementEntropyQuantum2004,brandaoAreaLawEntanglement2013,calabreseEntanglementEntropyConformal2009}. Moreover, it enables extraction of universal quantities such as the central charge in conformal field theory (CFT) \cite{calabreseEntanglementEntropyConformal2009,holzheyGeometricRenormalizedEntropy1994}. The degeneracy of the entanglement spectrum encodes information about edge states \cite{liEntanglementSpectrumGeneralization2008,pollmannEntanglementSpectrumTopological2010,zhouTopologicalEdgeStates2025,chandranBulkedgeCorrespondenceEntanglement2011,qiGeneralRelationshipEntanglement2012,yaoEntanglementEntropyEntanglement2010}. 
While the entanglement entropy in static systems has been extensively investigated \cite{vidalEntanglementQuantumCritical2003,gioevEntanglementEntropyFermions2006,fidkowskiEntanglementSpectrumTopological2010,swingleEntanglementEntropyFermi2010,aresEntanglementAsymmetryProbe2023,groverEntanglementInteractingFermions2013,yaoEntanglementEntropyEntanglement2010,wangDetectingEdgeDegeneracy2015,choiFinitetemperatureEntanglementNegativity2024}, its behavior in periodically driven systems hosting multiple Floquet edge modes remains much less explored. In contrast to static systems, periodic driving can induce diverse phases and transitions, including Floquet topological zero modes, $\pi$ modes, and hybrid phases hosting both zero and $\pi$ modes \cite{rechtsmanPhotonicFloquetTopological2013b,eckardtColloquiumAtomicQuantum2017a,rudnerAnomalousEdgeStates2013a,lindnerFloquetTopologicalInsulator2011a,jiGeneralizedBulkboundaryCorrespondence2024b,kitagawaTopologicalCharacterizationPeriodically2010,fruchartComplexClassesPeriodically2016a,heFloquetChernInsulators2019a}. In addition, experimental advances in ultracold atomic systems and phononic platforms have enabled direct measurement of entanglement related observables \cite{pichlerMeasurementProtocolEntanglement2016,chenExperimentalObservationClassical2019,islamMeasuringEntanglementEntropy2015,linMeasuringEntanglementEntropy2024}, providing new avenues to explore topological phenomena.

In recent years, entanglement entropy in non-Hermitian quantum systems has attracted increasing attention, offering a powerful framework for analyzing non-Hermitian topological phase transitions \cite{cheongManybodyDensityMatrices2004,herviouEntanglementSpectrumSymmetries2019,changEntanglementSpectrumEntropy2020a,guoEntanglementEntropyNonHermitian2021,chenEntanglementNonhermiticityDuality2021,okumaQuantumAnomalyNonHermitian2021a,chenQuantumEntanglementNonHermitian2022,ortega-tabernerPolarizationEntanglementSpectrum2022,kawabataEntanglementPhaseTransition2023,hsiehRelatingNonHermitianHermitian2023,legalVolumetoareaLawEntanglement2023,tuRenyiEntropiesNegative2022,yangEntanglementEntropyGeneralized2024,rottoliEntanglementHamiltonianNonHermitian2024,wangAbsenceMeasurementinducedEntanglement2024,zhouEntanglementPhaseTransitions2024b,zhangYangLeeEdgeSingularity2025}.
For non-Hermitian systems, the bulk eigenstates under open boundary conditions (OBC) tend to localize at the system boundaries, which is named the non-Hermitian skin effect (NHSE) \cite{yaoEdgeStatesTopological2018a,longhiProbingNonHermitianSkin2019a,songNonHermitianSkinEffect2019a,jiangInterplayNonHermitianSkin2019a,leeAnatomySkinModes2019a,liTopologicalEnergyBraiding2022b,liCriticalNonHermitianSkin2020a,jiNonHermitianSecondorderTopological2024a}. This effect renders the spectrum highly sensitive to boundary conditions, leading to distinct gap structures between periodic (PBC) and OBC \cite{yaoNonHermitianChernBands2018a,wanQuantumSqueezingInducedPointGapTopology2023a,jiFloquetEngineeringPointgapped2025a,caoNonHermitianBulkboundaryCorrespondence2021b,fuDegeneracyDefectivenessNonHermitian2022,guoExactSolutionNonHermitian2021a}. The NHSE also leads to the breakdown of the bulk–boundary correspondence (BBC) \cite{kawabataNonBlochBandTheory2020a,yokomizoNonBlochBandTheory2019a,yangNonHermitianBulkBoundaryCorrespondence2020,zhangBulkboundaryCorrespondenceNonHermitian2020a,liuFateMajoranaZero2021a,liExactSolutionsDisentangle2025,wangAmoebaFormulationNonBloch2024a,wangTheorySpectralSplitting2025}. While entanglement phase transitions have been observed in non-Hermitian systems \cite{couvreurEntanglementNonunitaryQuantum2017,bacsiDynamicsEntanglementExceptional2021,modakEigenstateEntanglementEntropy2021,leeExceptionalBoundStates2022,fossatiSymmetryresolvedEntanglementCritical2023,fengAbsenceLogarithmicAlgebraic2023,wangScalingLawsNonHermitian2023,oritoEntanglementDynamicsManybody2023,liEmergentEntanglementPhase2024}, the associated many-body correlations and critical behavior remain poorly understood in Floquet systems.

In this work, we study the Floquet SSH model with periodically modulated of the intracell hopping. By analyzing the entanglement entropy and the entanglement spectrum, we characterize topological phase transitions among topological zero modes, topological $\pi$ modes, and hybrid phase hosting both topological zero and $\pi$ modes. The central charge of the CFT is determined from the finite-size scaling of the entanglement entropy. In addition, within the hybrid phase hosting both topological zero and $\pi$ modes, we investigate the variation of the entanglement spectrum and the entanglement entropy $S$ with the driving frequency $\omega$. We further incorporate next-nearest-neighbor hopping and non-Hermitian terms to examine the robustness and generality of our approach. These results deepen the understanding of quantum criticality in Floquet non-Hermitian systems and establish a general framework for identifying topological phase transitions in non-Hermitian periodically driven systems.

\section{II. Model Hamiltonian}

We consider a one-dimensional Floquet Su–Schrieffer–Heeger (SSH) model subjected to periodic driving. For simplicity, only nearest-neighbor hopping is included. The time-dependent Bloch Hamiltonian reads \cite{suSolitonsPolyacetylene1979,caoNonHermitianBulkboundaryCorrespondence2021b}:
\begin{equation}
H(k,t)=[d_x+\lambda\cos(\omega t)]\sigma_x+d_y\sigma_y,
\end{equation}
where $d_{x}=t_{1}+t_{2}\cos k$, $d_{y}=t_{2}\sin k$, and $\sigma_{x,y}$ are the Pauli matrices. The parameters $t_{1}$ and $t_{2}$ denote the intracell and intercell hopping amplitudes, respectively. $\lambda$ characterizes the driving strength. The Hamiltonian is time-periodic, satisfying $H(k,t)=H(k,t+T)$, where $T=2\pi/\omega$ and $\omega$ is the frequency of the periodic driving. In the frequency domain \cite{lindnerFloquetTopologicalInsulator2011a,rudnerAnomalousEdgeStates2013a}, the Schrödinger equation can be written as:
\begin{equation}
\sum_{m^{\prime}}\mathcal{H}_{m,m^{\prime}}\left(k\right)|\psi_{n}^{\left(m^{\prime}\right)}\left(k\right)\rangle=\varepsilon_{n}\left(k\right)|\psi_{n}^{\left(m\right)}\left(k\right)\rangle.
\end{equation}
Here, the Floquet Hamiltonian  
$\mathcal{H}_{m,m^{\prime}}(k)=m\omega\delta_{m,m^{\prime}}\mathbf{I}+H_{m-m^{\prime}}(k)$ and
$H_{m}(k)=\frac{1}{T }\int_{0}^{T}dtH(k,t)\mathrm{exp}(-im\omega t)$. 

Owing to chiral symmetry, the Floquet Hamiltonian satisfies $\mathcal{C}^{-1}\mathcal{H}\mathcal{C}=-\mathcal{H}$, which enforces the quasienergy spectrum to appear in pairs $(E,-E)$\cite{caoNonHermitianBulkboundaryCorrespondence2021b}. 
In the weak driving case, the Floquet spectrum replicates the static one shifted by integer multiples of $\omega$. To avoid overcounting, we restrict the quasienergy range of the eigenstates to $\epsilon<|\omega/2|$.

\section{III. Entanglement entropy and spectrum}
The entanglement entropy provides a powerful diagnostic of topological phase transitions by capturing the critical structure of the many-body wavefunction $|\psi\rangle$. For the 1D Hermitian Floquet SSH model, the system can be divided into two subsystems, $A$ and $B$. The reduced density matrix of subsystem $A$ is obtained by tracing out the degrees of freedom of the remaining subsystem $B$ \cite{yangEntanglementEntropyGeneralized2024,zhouEntanglementPhaseTransitions2024a}:
\begin{equation}
    \rho_A=\mathrm{Tr}_B\left(|\psi\rangle\langle\psi|\right).
\end{equation}
The von-Neumann entanglement entropy is then defined as \cite{tuRenyiEntropiesNegative2022,yangEntanglementEntropyGeneralized2024,rottoliEntanglementHamiltonianNonHermitian2024}:
\begin{equation}
    S_A = -\mathrm{Tr} \left( \rho_A \ln \rho_A \right).
\end{equation}

For noninteracting fermionic systems, the entanglement entropy can be efficiently evaluated using the single-particle correlation matrix \cite{,cheongManybodyDensityMatrices2004,herviouEntanglementSpectrumSymmetries2019,changEntanglementSpectrumEntropy2020a,guoEntanglementEntropyNonHermitian2021}:
\begin{equation}
C_{ij}=\bra{\psi}\hat{c}_{i}^{\dagger}\hat{c}_{j}\ket{\psi},
\label{matrix}
\end{equation}
where the indices $i,j$ are restricted to sites within subsystem $A$. 
The eigenvalues $\{\xi_l\}$ of the correlation matrix constitutes the entanglement spectrum of subsystem $A$. In terms of these eigenvalue, the entanglement entropy between subsystem $A$ and the remaining part $B$ of the entire system can be expressed as \cite{tuRenyiEntropiesNegative2022,yangEntanglementEntropyGeneralized2024}:
\begin{equation}
S_A=-\sum_l[\xi_l\log(\xi_l)+(1-\xi_l)\log(1-\xi_l)].
\end{equation}

By analyzing the finite-size scaling of the entanglement entropy, we can extract universal information about quantum critical points \cite{leeManybodyApproachNonHermitian2020,zhouEntanglementPhaseTransitions2024a,chenCharacterizingBulkboundaryCorrespondence2022}. For a one-dimensional critical system described by conformal field theory (CFT), the entanglement entropy obeys the universal logarithmic scaling form \cite{pasqualecalabreseEntanglementEntropyQuantum2004,laflorencieQuantumEntanglementCondensed2016}:
\begin{equation}
S_A(N)=\frac{c}{3}\ln(N)+a,
\label{half}
\end{equation}
where $c$ is the central charge, $N$ denotes the total lattice length, and $a$ is a constant. In our calculations, the system is bipartitioned into two equal halves, with the subsystem size fixed as $L=N/2$, while the total lattice length $N$ is varied.

\section{IV. Phase diagram from entanglement entropy}

\begin{figure}
\centering
\includegraphics[width=0.48\textwidth]{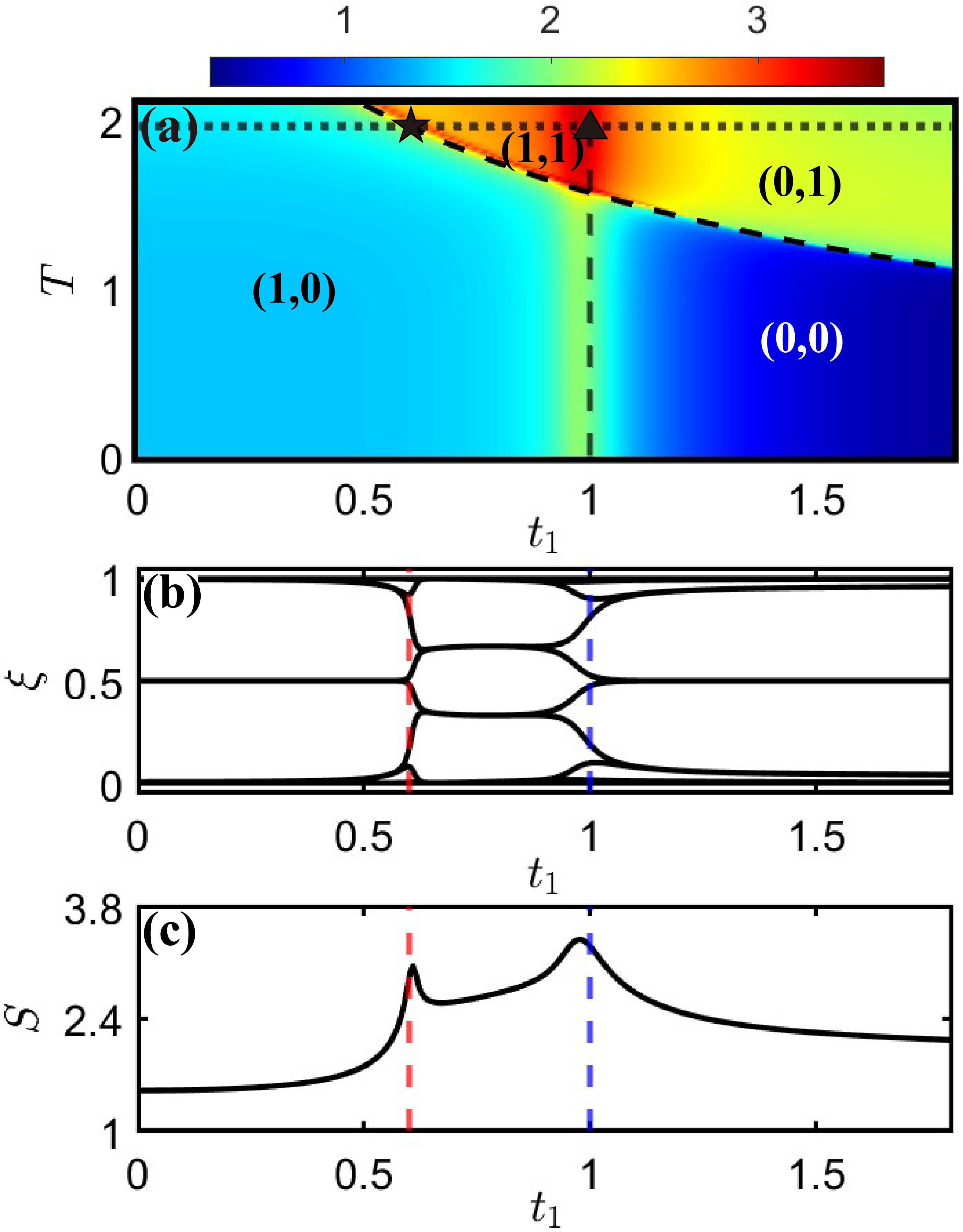}
 \caption{(a) Phase diagram of entanglement entropy in the $T-t_1$ plane, the colorbar represents the magnitude of the entanglement entropy. (b) and (c) The entanglement spectrum $\xi$ and entanglement entropy $S$ as functions of the $t_1$. The black pentagram and red dashed lines correspond to the emergence of topological $\pi$ modes, whereas the black triangle and blue dashed lines correspond to the disappearance of topological zero modes. The parameters are set as follows: $t_2=1$, $\lambda=0.8$, and $N=60$.
 } 
\label{Phase diagram} 
\end{figure}

In contrast to static systems, periodic driving can induce a distinct class of topological $\pi$ modes. By calculating the entanglement entropy of the system, we constructed the entanglement entropy phase diagram in the $T-t_1$ plane, as shown in Fig.~\ref{Phase diagram}(a). According to the magnitude of the entanglement entropy, the parameter space can be divided into four distinct regimes: the trivial phase with no edge modes $(0, 0)$, the phase hosting only topological zero modes $(1, 0)$, the phase hosting only topological $\pi$ modes $(0, 1)$, and the coexistence phase hosting both topological zero modes and $\pi$ modes $(1, 1)$. Specifically, we observe that the entanglement entropy $S$ increase with the number of topological edge modes. Under weak driving conditions, the phase boundaries can be obtained as:
$t_1=\pm\,{t_2}$, and $t_1=\pm\,{(\frac{\omega}{2}-t_2)}$.

Compared with phases hosting one type of edge modes, the entanglement entropy in the coexistence phase of topological zero and $\pi$ modes exhibits more intricate behavior. As indicated by the dashed line at $\omega = 3.2$ in Fig.~\ref{Phase diagram}(a), the system can be driven through three distinct topological phases by tuning $t_1$: (i) only having topological zero modes, (ii) having both topological zero and $\pi$ modes, (iii) only having topological $\pi$ modes. We show the entanglement spectrum as a function of the $t_1$ at $\omega = 3.2$ in Fig.~\ref{Phase diagram}(b). The red and blue dashed lines mark the analytically determined transition points associated with the $\pi$ and zero modes, respectively. In the topological zero modes phase, the entanglement spectrum exhibits a two-fold degeneracy at $\xi=0.5$, which is consistent with the energy spectrum features under OBC and reflects the two-fold degeneracy of zero modes. In the coexistence phase of topological zero and $\pi$ modes, their coupling induces a gap in the entanglement spectrum, which contrasts with the single-phase behavior observed in the topological zero modes phase and topological $\pi$ modes phase. As the system transitions to the $\pi$ modes phase, the gap in the entanglement spectrum closes again, restoring the two-fold degeneracy at $\xi = 0.5$. This reflects the disappearance of zero modes and the preservation of $\pi$ modes. The entanglement entropy as a function of the $t_1$ is presented in Fig.~\ref{Phase diagram}(c) for the case of $\omega = 3.2$. At the phase transition points, it exhibits sharp enhancements and forms pronounced peaks, thereby signaling topological phase transitions.

\begin{figure}
\centering
\includegraphics[width=0.48\textwidth]{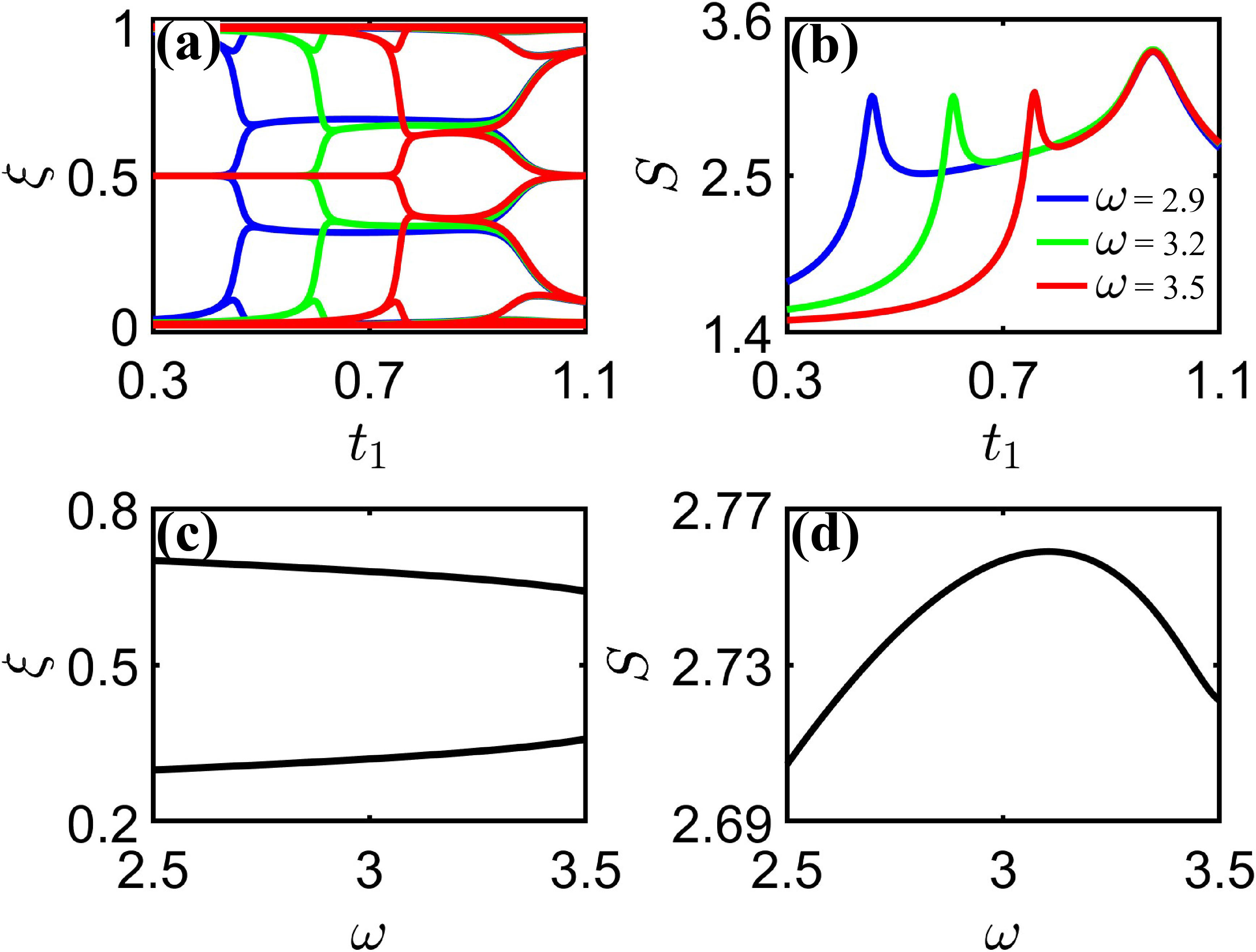}
 \caption{(a) The entanglement spectrum $\xi$ as a function of $t_1$. (b) The entanglement entropy $S$ as a function of $t_1$. In (a) and (b), curves with the same color correspond to the same driving frequency $\omega$. (c) The gap in the entanglement spectrum as a function of the $\omega$ at $t_1=0.8$ in the coexistence phase of topological zero and $\pi$ modes. (d) The entanglement entropy $S$ as a function of $\omega$ at $t_1=0.8$. Other parameters used in the calculations are set as follows: $t_2=1$, $\lambda=0.8$, and $N=60$.
 }
\label{omegaw}
\end{figure}

We have shown that the coupling between zero and $\pi$ modes splits the entanglement spectrum into two distinct two-fold degenerate branches. To explore this, We further investigate the entanglement spectrum $\xi$ as a function of $t_1$ for various $\omega$, as shown in Fig.~\ref{omegaw}(a). As $t_1$ increases, the twofold degeneracy of the entanglement spectrum at $\xi=0.5$ is lifted and a gap opens. At higher $t_1$, the gap closes and the twofold degeneracy reappears. The alteration in the entanglement spectrum structure signals the phase transition in the system and corresponds to the emergence or disappearance of zero or $\pi$ topological edge modes. Fig.~\ref{omegaw}(b) shows the $S$ as a function of $t_1$ for various $\omega$. As the driving frequency $\omega$ varies, the $\pi$ mode transition point exhibits a pronounced shift, demonstrating that $\omega$ tunes the phase boundaries. Figure ~\ref{omegaw}(c) presents the splitting of the entanglement spectrum versus driving frequency $\omega$ in the coexistence regime of zero and $\pi$ modes. The entanglement spectrum gap is tuned by the periodic driving frequency $\omega$. With increasing $\omega$, the gap of entanglement spectrum decreases. The entanglement entropy $S$ as a function of $\omega$ at $t_1=0.8$, as shown in Fig.~\ref{omegaw}(d). 

\section{V. Entanglement spectrum and entropy under half separation}

\begin{figure}
\centering
\includegraphics[width=0.48\textwidth]{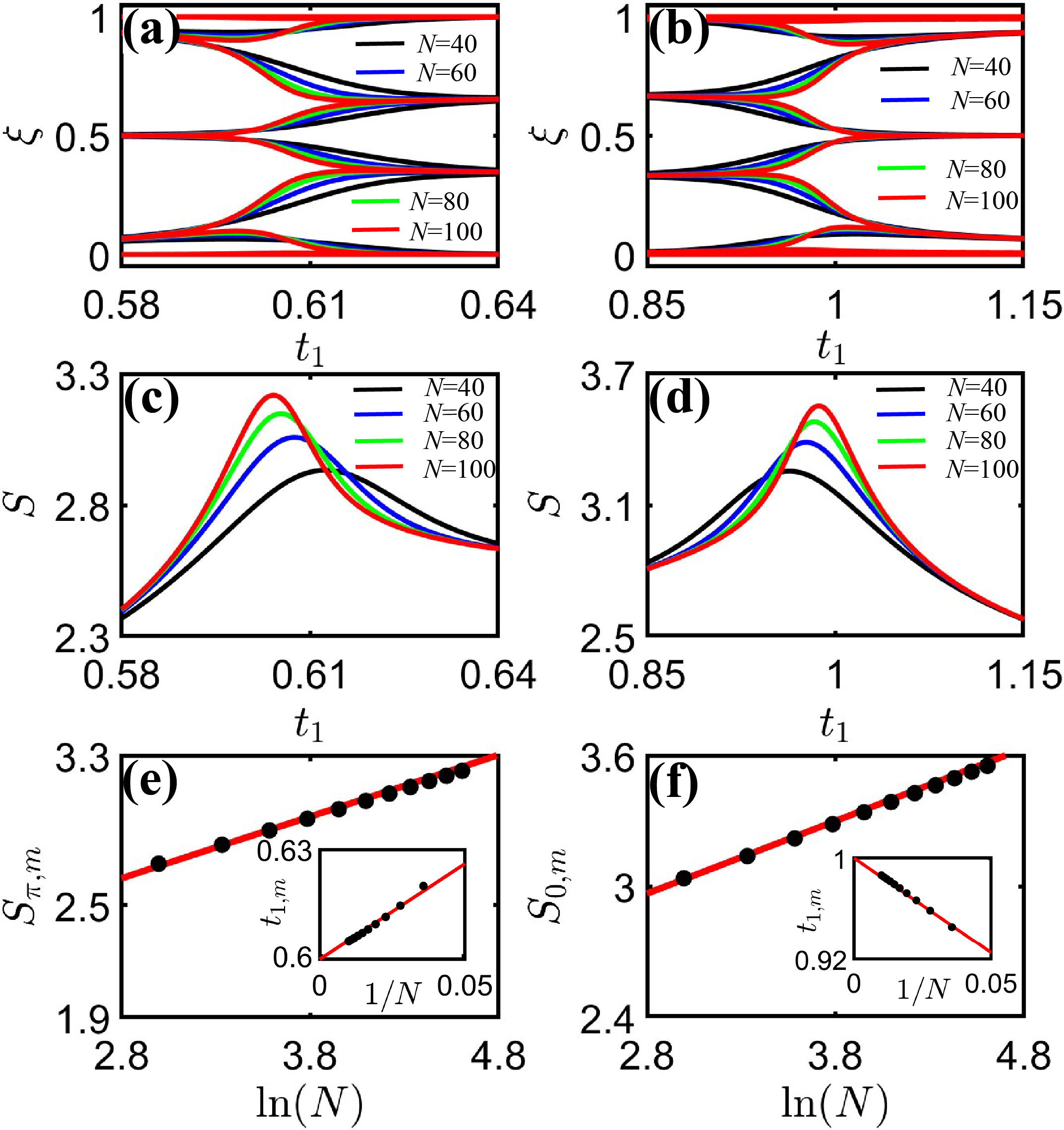}
\caption{(a) and (b) The entanglement spectrum $\xi$ as a function of $t_1$, shown near the transition points of the topological $\pi$ mode and zero mode, respectively. (c) and (d) Entanglement entropy as a function of $t_1$ in the vicinity of the phase transition points associated with the $\pi$ mode and the zero mode, respectively. (e) and (f) The logarithmic scaling of entanglement entropy with the total lattice size $N$ at the transition points of the topological zero and $\pi$ modes, respectively. The inserts show the finite-size scaling of $t_{1,m}$ at the maximum of the entanglement entropy $S_{\pi}$ and $S_0$, respectively.
Other parameters are set as follows: $t_2=1$, $\lambda=0.8$, and $\omega=3.2$.
} 
\label{Nscaling} 
\end{figure}

Both the entanglement entropy and entanglement spectrum exhibit size-dependent effects. To explore this, we focus on the transition points of the $\pi$ mode and zero mode transitions and compute the entanglement spectra under equal bipartition of varying system sizes to identify the topological phase transitions, as shown in Figs.~\ref{Nscaling}(a) and~\ref{Nscaling}(b). In the coexistence phase of topological zero and $\pi$ modes, their coupling induces a gap in the entanglement spectrum near $\xi=0.5$, which contrasts with the single-phase behavior observed in the topological zero modes phase and topological $\pi$ modes phase. The entanglement spectrum gap opening does not involve the zero modes and $\pi$ modes merging with the bulk. Rather, we still have edge modes, but the nature of the edge modes is different when topological zero modes and $\pi$ modes couple to each other. 

Following the analysis of the topological phase transitions through the entanglement spectrum, we investigate the entanglement entropy under half-separation for different system sizes. As the parameter $t_1$ varies, significant changes in entanglement entropy can be observed near the transition point, with a peak shown in Figs.~\ref{Nscaling}(c) and~\ref{Nscaling}(d). Calculations for different system sizes show that the peak of the entanglement entropy deviates slightly from the analytical value, which can be attributed to finite-size effects. As the system size increases, the peak value of the entanglement entropy grows, and $t_{1,m}$ approaches the analytical value. 

We further study the finite-size scaling of the entanglement entropy under half separation. As shown in Fig~\ref{Nscaling}(e), the scaling behavior at the critical point separating the topological zero modes phase and the coexistence phase follows the logarithmic law. By fitting the numerical data to the formula $S=\tfrac{c}{3}\ln N+ a$, we obtain the central charge of $c=1$. As shown in Fig.~\ref{Nscaling}(f), the scaling behavior of the entanglement entropy at the transition point between the coexistence phase and the topological $\pi$ modes phase also follows the logarithmic scaling law and $c=1$. The finite-size scaling of the $t_1$ with respect to the $N$ at the transition points of the $\pi$ modes and zero modes are shown in the insert of Figs~\ref{Nscaling}(c) and~\ref{Nscaling}(d), respectively. As the total lattice size $N$ increases, the fitted phase transition points for the $\pi$ mode and the zero mode converge toward their respective analytical values $t_{1,m}=0.6$ and $t_{1,m}=1$, respectively.


\section{VI. Entanglement entropy in non-Hermitian system}

The entanglement entropy remains applicable in non-Hermitian systems and is insensitive to the specific form of the hopping. To confirm the universality of this approach, we further investigate the model with the non-Hermitian strength and next-nearest-neighbor hopping. Its Bloch Hamiltonian can be written in the following form \cite{yaoEdgeStatesTopological2018a,caoNonHermitianBulkboundaryCorrespondence2021b}:
\begin{equation}
H(k,t)=[d_x+\lambda\cos(\omega t)]\sigma_x+[d_y+i\gamma]\sigma_y,
\end{equation}
here, $d_{x}=t_{1}+(t_{2}+t_{3})\cos k$, $d_{y}=(t_{2}-t_{3})\sin k$, and $\sigma_{x,y}$ are the Pauli matrices. The parameters $t_{3}$ and $\gamma$ denote the next-nearest-neighbor hopping and non-Hermitian strength, respectively.

The concept of entanglement entropy can be extended to non-Hermitian systems. In the framework of biorthogonal quantum mechanics, the Hamiltonian has two types of eigenvectors \cite{brodyBiorthogonalQuantumMechanics2014,chenCharacterizingBulkboundaryCorrespondence2022,sunBiorthogonalQuantumCriticality2022,zhouEntanglementSpectrumEntropy2022}. 
\begin{equation}
    H |R_n\rangle = E_n |R_n\rangle, \quad H^\dagger |L_n\rangle = E_n^* |L_n\rangle,
\end{equation}
where $|L_n\rangle$ and $|R_n\rangle$ refer to the left and right eigenvectors, respectively, and they satisfy the biorthogonal condition $\langle L_m |R_n \rangle = \delta_{mn}$.
When the energy of non-Hermitian systems is complex, there are different ways to define the ground state. In our numerical calculations, we construct the Floquet ground state based on the real part of the energy \cite{guoEntanglementEntropyNonHermitian2021,chenCharacterizingBulkboundaryCorrespondence2022}.
\begin{equation}|\psi\rangle=\prod_{occ}\hat{c}_{n}^{\dagger}|0\rangle.
\end{equation}
$\hat{c}_{n}^{\dagger}$ is the creation operator at the $n$-th site. The biorthogonal density matrix $\rho = |\psi_L \rangle \langle \psi_R|$ is constructed using the left and right eigenstates. 
The entanglement entropy in non-Hermitian system between subsystem $A$ and the remaining part $B$ of the entire system can be expressed as \cite{tuRenyiEntropiesNegative2022,yangEntanglementEntropyGeneralized2024}:

\begin{equation}
S_A=-\sum_l(\xi_l\log(|\xi_l|)+(1-\xi_l)\log(|1-\xi_l|)),
\end{equation}
where $\xi_l$ denotes the set of eigenvalues of the correlation matrix in Eq.~\ref{matrix}.
\begin{figure}
\includegraphics[width=0.48\textwidth]{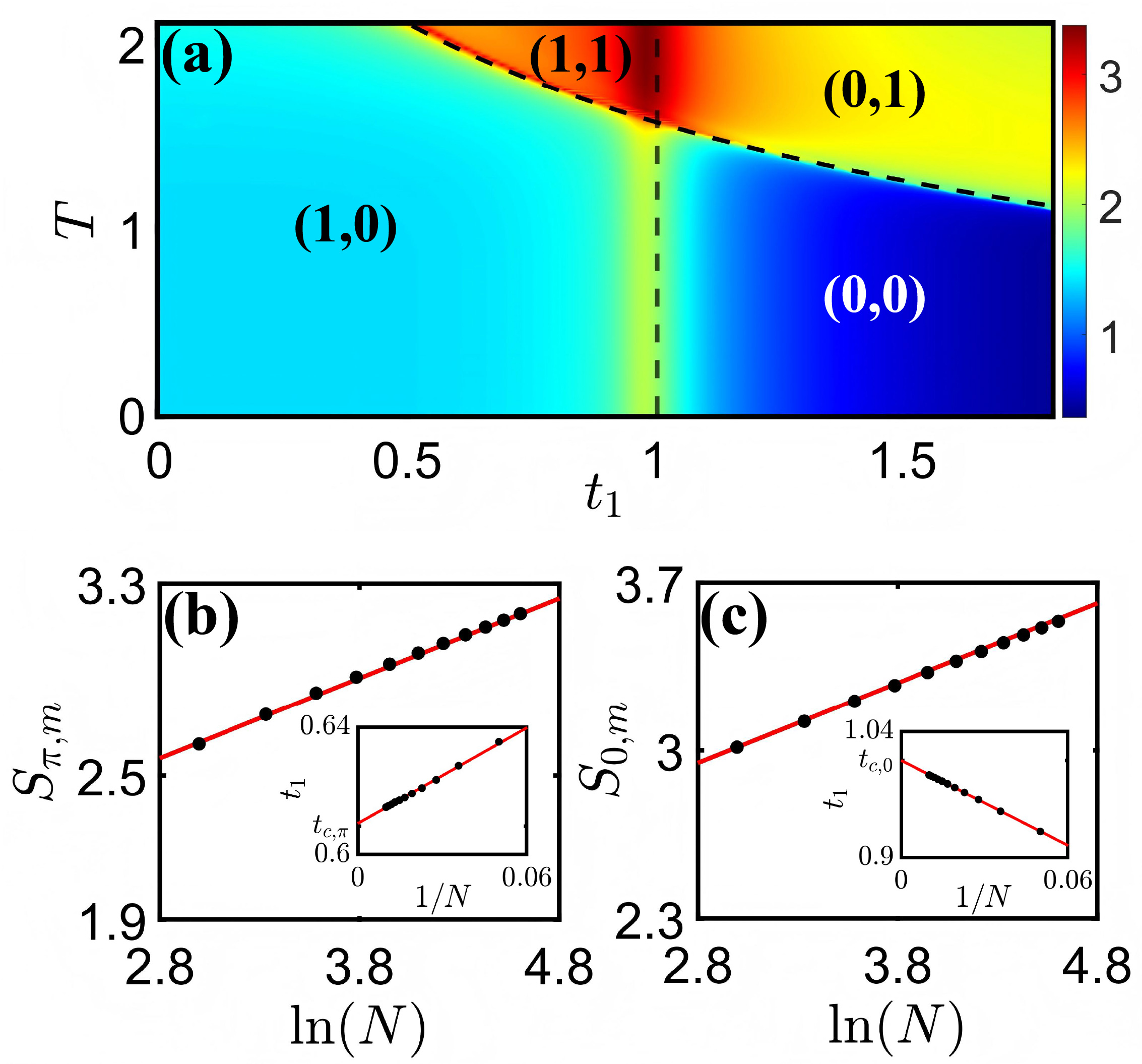}
\caption{(a) The entanglement entropy phase diagram of the non-Hermitian system in $T-t_1$ plane with $\gamma=0.1$ and $N=60$. (b) and (c) The logarithmic scaling of entanglement entropy with the total lattice size $N$ at the transition points of the topological zero and $\pi$ modes, respectively. The inserts show the finite-size scaling of $t_1$ at the maximum of the entanglement entropy $S_{\pi}$ and $S_0$, respectively. Other parameters used in the calculations are set as follows: $t_2=1$, $t_3=0$, $\lambda=0.8$, $\omega=3.2$, and $\gamma=0.1$.} 
\label{nh}
\end{figure}
 
We first investigate the entanglement entropy $S$ with the non-Hermitian strength $\gamma=0.1$ and $t_{3}=0$.  In this case, the phase boundaries are modified to:
$t_1=\pm{\sqrt{{t_2}^2+{\gamma}^2}}$ and $t_1=\pm{\sqrt{(\frac{\omega}{2}-t_2)^2+\gamma^2}}$. When the energy of non-Hermitian systems is complex, we construct the ground state by half-filling according to the real part of the energy. Figure~\ref{nh}(a) shows the entanglement entropy phase diagram of the non-Hermitian system in $T-t_1$ plane with $\gamma=0.1$, and the black dashed lines show the analytical phase boundary. In different phase regimes, the entanglement entropy increases with the number of topological edge modes. The entanglement entropy exhibits a peak at the phase boundary, reflecting the topological phase transition occurring in the system. 

Within each topological phase, the entanglement entropy grows with the number of topological edge modes, which is consistent with the behavior observed in Hermitian systems. However, the phase transition points of the non-Hermitian system deviate significantly from its Bloch case. This discrepancy stems from the breakdown of the conventional bulk-boundary correspondence. The conventional Bloch band theory based on Brillouin zone cannot accurately describe the topological and boundary phenomena of non-Hermitian systems.

\begin{figure}
\includegraphics[width=0.48\textwidth]{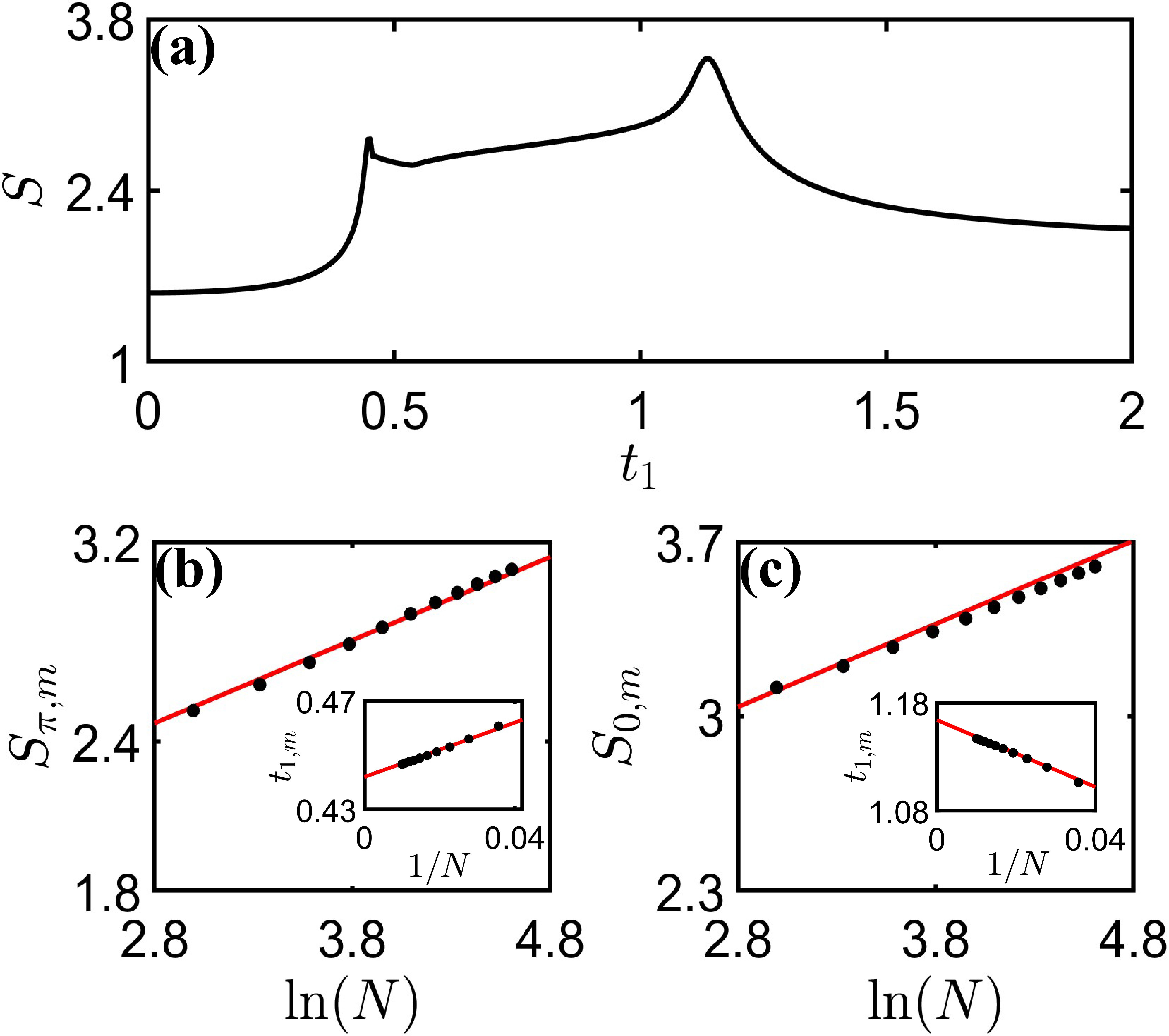}
\caption{(a) The entanglement entropy $S$ as a function of the $t_1$. (b) and (c) The logarithmic scaling of entanglement entropy with the total lattice size $N$ at the transition points of the topological zero and $\pi$ modes, respectively. The inserts show the finite-size scaling of $t_{1,m}$ at the maximum of the entanglement entropy $S_{\pi}$ and $S_0$, respectively. Other parameters used in the calculations are set as follows: $t_2=1$, $t_3=0.15$, $\lambda=0.8$, $\omega=3.2$, and $\gamma=0.1$.} 
\label{t3}
\end{figure}

The entanglement entropy faithfully captures the actual phase transitions in non-Hermitian systems. To further illustrate this point, we show the finite-size scaling of the  entanglement entropy at the $\pi$ mode and zero mode phase transitions in Figs.~\ref{nh}(b) and~\ref{nh}(c), respectively. As the system size $N$ increases, the entanglement entropy at both transition points exhibits logarithmic scaling. The central charges $c=1$ can be obtained by fitting the entropy to the scaling formulas given in Eq.~\ref{half}, the entropy in both cases follows similar logarithmic scaling. Notably, the phase transition points in non-Hermitian systems differ fundamentally from those in Hermitian systems. The insets of Figs.~\ref{nh}(b) and~\ref{nh}(c) show that, as the system size $N$ increases, the fitted phase transition points for the $\pi$ mode and the zero mode converge toward their respective analytical values $t_{c,\pi}=0.608$ and $t_{c,0}=1.005$, respectively.

As displayed in Fig.~\ref{t3}(a), we present the behavior of the entanglement entropy as a function of $t_1$ when both the non-Hermitian parameter $\gamma$ and the next-nearest-neighbor hopping $t_3$ are included. Even though the introduction of $t_3$ renders the analytical determination of the transition points intractable, the entanglement entropy still exhibits sharp peaks at the phase boundaries, enabling a clear identification of the topological $\pi$ mode and zero mode transitions. To further confirm the critical behavior of the peaks, we perform a finite-size scaling analysis of the entanglement entropy, as shown in Figs.~\ref{t3}(b) and~\ref{t3}(c). For both the $\pi$ mode and zero mode transitions, the entanglement entropy follows the logarithmic scaling formula given in Eq.~\ref{half}, and the fitted central charge is $c=1$. The inserts of Figs.~\ref{t3}(b) and \ref{t3}(c) show the finite-size scaling of $t_{1,m}$ with respect to $N$ at the phase transition points for the $\pi$ mode and the zero mode, respectively. 

\section{VII. Summary and discussion}
In this work, we have investigated the topological phase transitions of the Floquet non-Hermitian SSH model using the entanglement entropy and the entanglement spectrum. The entanglement spectrum exhibits characteristic degeneracy patterns that faithfully reflect the edge mode structure. In phases hosting a single type of edge mode, the entanglement spectrum exhibits the expected two-fold degeneracy at $\xi=0.5$. In contrast, the entanglement spectrum undergoes splitting in the coexistence phase, reflecting the coupling between topological zero and $\pi$ modes. Furthermore, the entanglement entropy displays pronounced peaks at the topological transition points and follows the logarithmic scaling with system size. 
We further examine the the influence of the non-Hermitian term $\gamma$ and next-nearest-neighbor hopping $t_3$. Although these terms considerably complicate the analytic determination of the transition boundaries, the entanglement entropy still exhibits clear peaks that accurately locate the topological phase transitions. Finite-size scaling confirms that the peak positions approach the true critical points as the system size increases. This shows that the entanglement entropy remains a robust and universal diagnostic tool in non-Hermitian Floquet systems. 

Our work promotes the understanding of entanglement properties in general non-Hermitian systems. The entanglement behavior of systems with complex spectral characteristics, including gapless bands \cite{yuUniversalEntanglementSpectrum2024,zhongQuantumEntanglementFermionic2025,banerjeeEntanglementSpectrumGapless2025}, band braiding \cite{zhangEntanglementManifestationKnot2025}, and exceptional points \cite{leeExceptionalBoundStates2022,yangEntanglementEntropyGeneralized2024}, is expected to exhibit rich physics and merits systematic investigation. Moreover, the entanglement entropy defined on the generalized Brillouin zone merits further investigation. Progress in these areas will deepen the understanding of quantum critical behavior in Floquet non-Hermitian systems, and will help establish entanglement entropy as a universal diagnostic applicable to both Hermitian and non-Hermitian systems.

\section{Acknowledgments}
This work is supported by Natural Science Foundation of Jiangsu Province (Grant No. BK20231320).

\bibliography{entropyreference}
\end{document}